\documentclass[aps,pra,twocolumn,amsmath,amssymb,showpacs]{revtex4}
\usepackage{graphicx}
\usepackage{dcolumn}
\usepackage{bm}
\input epsf

\begin{document}

\preprint{}

\title{Prospects for an electron electric dipole moment search in metastable ThO and ThF$^{\rm +}$}

\author{Edmund R. Meyer}
\email{meyere@murphy.colorado.edu}
\affiliation{JILA, NIST and University of Colorado, Department of Physics,
  Boulder, Colorado 80309-0440, USA}
\author{John L. Bohn}
\affiliation{JILA, NIST and University of Colorado, Department of Physics,
  Boulder, Colorado 80309-0440, USA}

\date{\today}

\begin{abstract}
The observation of an electron electric dipole moment (eEDM) would
have major ramifications for the standard model of physics. Polar molecules 
offer a near-ideal laboratory for such searches due to the large effective 
electric field (${\bf F}_{\rm eff}$), on order of tens of GV/cm that can be 
easily oriented in the lab frame. We present an improved method for simply 
and accurately determining ${\bf F}_{\rm eff}$, in a heavy polar molecule, 
allowing for a quick determination of candidates for an eEDM experiment. 
We apply this method to ThO and ThF$^{\rm +}$, both of which possess 
metastable $^3\Delta$ electronic states. The values of ${\bf F}_{\rm eff}$ 
in ThO and ThF$^{\rm +}$ are estimated to be 104\,GV/cm and 90\,GV/cm respectively, 
and are therefore two of the best known candidates for the eEDM search.
\end{abstract}


\maketitle

One of the more spectacular goals of in the rapidly growing field of 
ultra-cold molecules\cite{QuoVadis} is the search for the electric
dipole moment of the electron (eEDM).  If found, the eEDM would be a
touchstone against which ideas beyond the Standard Model(SM), especially 
Supersymmetry, could be checked. On the other hand, if it is not found but 
experiments can push the eEDM's value below its current experimental
limit of $|d_e| < 1.6 \times 10^{-27}$\,e-cm\cite{Commins}, then key 
supersymmetric extensions to the SM become much less likely to be true\cite{barr}. 
Thus, table-top experiments at the lowest achievable energies are a direct 
compliment to collider experiments at the highest. 

The current experimental limit on the eEDM originates from a
high-resolution spectroscopic experiment on atomic thallium
\cite{Commins}. The ability to make this kind of precise measurement in 
atoms stems from key insights by Sandars\cite{Sandars} and contributions 
by Flambaum and Khriplovich\cite{khrip1,flam}. These works show that 
the {\it apparent} electric field acting on an
electron inside an atom can be far larger than the field that is
directly applied by a macroscopic laboratory apparatus.  A further
insight by Sandars\cite{Sandarsmol} pointed out that, for an electron
inside a polar molecule, the effective electric field ${\bf F}_{\rm
eff}$ can be even larger.  This field is enhanced by relativistic
effects and scales with nuclear charge $Z$ as $Z^3$, therefore 
preferring heavy elements, the same as in the atomic case. 
Effective electric fields as large as tens of GV/cm 
have been anticipated in certain heavy molecules, the largest so 
far being in HgF where ${\bf F}_{\rm eff} \sim 99$\;GV/cm\cite{pbf}.

In practice, ${\bf F}_{\rm eff}$ is manifested in the
molecule by its ability to distort an electron orbital of nominal
$s$ symmetry on the heavy atom into a combination of an $s$- 
and $p$-orbital. For this reason, several candidate molecules are
composed of a heavy $(n\,s^2){}^1S$-state atom with fluorine -- which is adept at
drawing electrons toward itself -- therefore accounting for the needed
distortion. Molecules such as BaF\cite{baf}, YbF\cite{ybf}, HgF 
and PbF\cite{pbf} have all been considered.  Departing from this trend,
DeMille {\it et al.} proposed the metastable $a(1)$ state 
of PbO\cite{pbo}. More recently, an alternative experimental effort has 
proposed to trap molecular ions, counting on the long trap lifetimes to 
improve the signal\cite{Stutz}. In response, several ions were proposed,
including HI$^+$ \cite{Isaev}, PtH$^+$, HfH$^+$\cite{meyerbohn07}, 
and HfF$^+$\cite{meyerbohn07,Petrov}.

The molecular ion experiment has stressed the desirability of
molecules that are easily polarized in small laboratory electric
fields.  For ion traps in particular, this attribute is a
necessity because strong electric fields would ruin the trap's
characteristics. Molecules with this property that also have a reasonable 
value of ${\bf F}_{\rm eff}$, are of $({\rm s}\sigma {\rm d}\delta)^3\Delta$ 
electronic symmetry. In such a molecule, the overall $\Delta$ symmetry
guarantees a small $\Omega$-doubling, hence easy polarizability,
while the electron in the $\sigma$-orbital may still contribute a
large ${\bf F}_{\rm eff}$\cite{meyerbohn07}. In addition, the magnetic $g$-factor 
of a ${}^3\Delta_1$ molecule is very small -- a tiny fraction of a 
Bohr magneton because $g_J\sim(g_L\Lambda-g_S\Sigma)$. This is zero, apart 
from diamagnetic and radiative corrections\cite{bnc}, providing less 
sensitivity to magnetic field noise.

The ACME collaboration\cite{acme} has seized on the potential utility of
$^3\Delta$ molecules in designing a new molecular beam eEDM
experiment. This experiment will use ThO molecules, which
possess a metastable $^3\Delta$ state not far above the ground
state.  This molecule is also attractive inasmuch as several of its isotopomers
have no nuclear spin, eliminating the complexities of hyperfine structure. 
Similarly, the JILA EDM team is interested in the isoelectronic cation 
ThF$^{\rm +}$, which is expected to be similar to ThO in its electronic 
structure, for use in an ion trap experiment\cite{eac}.

Until now, the most reliable estimates of ${\bf F}_{\rm eff}$ were determined 
using elaborate relativistic many-electron calculations. In this Letter we 
instead develop a competitive method based on nonrelativistic molecular 
structure calculations perturbed by the Hamiltonian arising from the eEDM. 
The method, based on an initial approximation in Ref. \cite{meyerbohn07}, is 
both fast an accurate. Indeed, we now reproduce, to within 25\% the vales 
${\bf F}_{\rm eff}$ for all species that have been treated by more elaborate 
relativistic theory. This circumstance opens the door for researchers to perform 
broad surveys of potential eEDM searches, and to evaluate the expected 
sensitivity of these experiments. As examples, we present in the Letter estimates 
of ${\bf F}_{\rm eff}$ for ThO and ThF$^{\rm +}$, finding them both to be excellent 
candidates.

The simple idea behind the method is that the presence of the eEDM causes 
a perturbation to the molecule's structure due to a relativistic 
effect\cite{khrip}. In Dirac notation the perturbation takes the form\cite{khrip1}
\begin{equation}
\label{EDM_Hamiltonian} H_{\rm d} = \left( \begin{array}{cc} 0 & 0\\
0 & 2d_e {\bf \sigma} \cdot {\bf F} \end{array} \right).
\end{equation}
Here ${\bf \sigma}$ represents the electron's spin, assumed to point in the same
direction as the eEDM; ${\bf F}$ is the local electric field
experienced by the electron inside the molecule, which is well
approximated by the Coulomb field due to the large nucleus; and
$d_e$ is the magnitude of the eEDM.

The influence of this interaction on the molecular spectrum is
computed in perturbation theory.  To do this, Ref.
\cite{meyerbohn07} expanded the molecular orbital into atomic orbitals,
\begin{equation}
\label{molecular_wave_function} |\Psi_{\rm mol} \rangle = \epsilon_s
|s \rangle + \epsilon_p |p \rangle + \cdots .
\end{equation}
Here the coefficients $\epsilon_s$ and $\epsilon_p$ represent the
contributions to $|\Psi_{\rm mol} \rangle$ due to the $s$ and $p$
atomic orbitals of the heavy atom.  Since the influence of the eEDM
relies on relativity, its contribution to the spectrum of the
molecule is dominated by motion of the electron near the heavy
nucleus.  Thus the $|s \rangle$ orbital is essential.  Moreover,
since the eEDM Hamiltonian \eqref{EDM_Hamiltonian} has odd symmetry
under parity, the matrix element $\langle s | H_d | s \rangle$
vanishes, and we must consider mixing with the heavy atom's $|p
\rangle$ orbital. The other atomic orbitals that comprise
$|\Psi_{\rm mol}\rangle$ are far less significant to our purposes.

The energy shift due to the perturbation is therefore
\begin{eqnarray}
\label{real_molecule_shift} \Delta E = \langle \Psi_{\rm mol} | H_d
| \Psi_{\rm mol} \rangle \approx 2 \epsilon_s \epsilon_p \langle s |
H_d | p \rangle.
\end{eqnarray}
To evaluate $\Delta E$, we estimate $\epsilon_{s,p}$ 
using nonrelativistic molecular structure software.  The
relativistic effects occur mostly in the matrix element $\langle s |
H_d | p \rangle$.  To evaluate this, we follow Ref. \cite{khrip} and
use one-electron Dirac-Coulomb wave functions in place of the true
orbitals $|s \rangle$ and $|p \rangle$.  This allows the integrals
to be done analytically.  Moreover, they are strikingly accurate
provided that the quantum numbers $n_{s,p}$ are replaced in
the result by the effective quantum numbers $\nu_{s,p}$, to
reflect the fact that these electrons actually exist around a
structured ionic core, as opposed to a bare Coulomb potential.

Using this approximation, the energy shift is\cite{meyerbohn07}
\begin{equation}
  \label{molecule_shift} \Delta E = |{\bf F}_{\rm eff}| d_e \approx
  \left[ -\frac{4 \sigma}{\sqrt{3}}
    h_0 \chi_0 \epsilon_s \epsilon_p \Gamma_{\rm rel} \frac{Z e}{a_0^2} \right] 
  d_e,
\end{equation}
which identifies the item in square brackets as the effective
electric field. In this expression, $\sigma$ is the projection of
the $s$-electron's spin onto the permanent dipole moment of the molecule; 
$e$ is the electron's charge; and $a_0$ is the Bohr radius, which 
means that the factor $e/a_0^2$ identifies a characteristic electric 
field strength. The relativistic factor, $\Gamma_{\rm rel}$, accounts 
for the overlap of the Dirac Coulomb functions:
\begin{equation}
  \label{relativistic_factor}
  \Gamma_{\rm rel} = -\frac{4 (Z \alpha)^2 Z_{\rm eff}^2}
	{\gamma (4 \gamma^2 - 1)  (\nu_s \nu_p)^{3/2}},
\end{equation}
where $\gamma = \sqrt{ (j+1/2)^2-(Z \alpha)^2 }$ is a 
familiar dimensionless quantity, and $Z$ is the atomic
number of the heavy nucleus.  By contrast, the effective charge
$Z_{\rm eff}$ is the charge of the heavy ion that the electron
orbits.  For example, in YbF, the bond is very nearly polar, and the
valence electron contributed by the Yb$^+$ orbits about a Yb-core with
$Z_{\rm eff} = 2$.  Further, in Eq. \eqref{molecule_shift}, $h_0$
is a coefficient (usually slightly less than unity) that accounts
for the fact that $|\Psi_{\rm mol} \rangle$ is typically a
combination of several different orbital configurations; and
$\chi_0$ is a further reduction of order unity that accounts for
spin-orbit mixing of $|\Psi_{\rm mol} \rangle $ with other
symmetries.  Heavy diatomic molecules are often written in a basis where 
only the projection of the total angular momentum $\Omega$ is a good 
quantum number. For example, the $^3\Delta_1$ electronic level can have 
some admixture of a nearby $^3\Pi_1$ electronic level.

The key to the success of the method is that the coefficients
$\epsilon_{s,p}$ can be extracted from nonrelativistic
molecular structure software (specifically, we have used the {\sc molpro}
suite of codes\cite{MOLPRO}). This is done via the overlap
integrals
\begin{equation}\label{overlap}
  \epsilon_{s,p} = \langle\Psi_{\rm mol}|\,(s,p)\rangle,
\end{equation}
where in Ref.\cite{meyerbohn07} the wave functions $|\Psi_{\rm mol} \rangle$ 
and $|(s,p) \rangle$ were obtained from {\it separate} calculations, one for the 
molecule and one for the heavy atom.  The atomic calculation moreover
identified orbitals with a particular principal quantum number $n$,
associated with the atom's valence shell.  In the context of
distorting the heavy atom's $s$ orbital, the appropriate $p$ orbital
must be the component $p_z$, i.e., the projection with vanishing
angular momentum about the molecular axis. 

The true molecular orbital is a combination of many atomic 
orbitals. Therefore, the $s$-component is not necessarily a state 
purely composed of a principal quantum number $n$, though it is 
likely dominated by one of them. The presence of 
the light atom will distort the atomic orbitals of the heavy atom 
and cause mixing of not only $s$ and $p$ but of $n$ and $n'$, 
where $n'$ is another atomic energy level of the heavy atom. 
These other configurations will contribute to a science signal as 
well, albeit somewhat weakly. 

Yet, a basis expansion of $|\psi_{\rm mol}\rangle$ can be done into any set of 
functions; for example, the Gaussian atomic basis set itself can be used. 
We can read off the contribution of each $s$ and $p$ heavy atom Gaussian 
to the molecular wavefunction. These are the functions that are variationally 
optimized to produce the molecular wavefunction and therefore utilize all 
the $s$ and $p$ character that is available to minimize the energy of the 
molecular state of interest.

Therefore, it is worthwhile (and more direct) to
extract the coefficients $\epsilon_{s,p}$ from the
molecular orbital calculation itself.  A molecular
orbital is comprised of a linear combination of Gaussian basis
functions with the same projection of $\lambda$. In the
case of the $\sigma$-molecular orbital, we can write out the
expansion of the molecular orbital as
\begin{eqnarray}\nonumber
\label{orbital_expansion}
  |\Psi_{\rm mol}\rangle &=& \sum_{i_s} c_{i_s}^{h}|s\sigma_{i_s}^{h}\rangle +
  \sum_{i_p}d_{i_p}^{h} |p\sigma_{i_p}^{h}\rangle +\cdots \\
  && +\sum_{j_s} c_{j_s}^{l}|s\sigma_{j_s}^{l}\rangle +
  \sum_{j_p}d_{j_p}^{l} |p\sigma_{j_p}^{l}\rangle +\cdots.
\end{eqnarray}
In this expression each ket represents, not an atomic orbital, but
rather a Gaussian basis function. The coefficients are therefore the
direct numerical output of the {\it ab initio} {\sc molpro} calculation.
The superscripts $h$ ($l$) identify functions centered on the heavy (light)
atom. Each Gaussian function is moreover 
identified by whether it has $s$ or $p$ (or $d$, $f$, etc.) symmetry
with respect to its atom. Viewed in this way, the net $s$-wave heavy atom 
character of $| \Psi_{\rm mol} \rangle$ is given by the sum of all
its $s$-wave contributions, meaning that
\begin{equation}
  \epsilon_s = 
  \frac{\sum_{k_s} c_{k_s}^h\langle \Psi_{\rm mol}|s\sigma_{k_s}^h\rangle}
       {\sum_{k_s\,l_s}c_{k_s}^{h\star}c_{l_s}^h\langle s\sigma_{k_s}^h
         |s\sigma_{l_s}^h\rangle}.
\end{equation}
Here the denominator is needed because the individual Gaussian
orbitals are not necessarily orthogonal to each other. A similar
expression holds for $\epsilon_p$.

This method incorporates the influence of several $n$'s, including those higher
than the nominal valence orbital, which may be mixed in virtually.
This advantage is also a drawback, however, because the 
lack of identification of a particular principal quantum number $n$ 
leads to an uncertainty in identifying the effective
quantum numbers $\nu_{s,p}$ in \eqref{relativistic_factor}. 

This same ambiguity was actually present in the previous form of the 
calculation\cite{meyerbohn07}. The atomic orbitals $|(s,p)\rangle$ from 
\eqref{overlap} merely represent the choice of a basis set. The basis 
could easily have been the neutral atom or the ion. Each choice of expansion 
would require a different choice of $\nu_{s,p}$. Because the original 
Gaussian basis set is optimized for the neutral atom, we extract 
$\nu_{s,p}$ from the neutral atom spectra. $\nu_{s,p}$ are related 
to the atomic quantum defect $\mu_{s,p}$ through\cite{fano76}
\begin{equation}\label{qd}
 \nu_l = n-\mu_l =  \sqrt{\frac{{\rm Ry}}{E_n}},
\end{equation}
where Ry is the Rydberg constant, and $E_n$ is the ionization energy of the 
heavy atom from the $|n\,l\rangle$ energy level. This method assumes 
that the spectrum for the atom follows a Rydberg like series for higher-$n$ 
levels. We then find the effective quantum number that arises from 
the structured core of the heavy atom. For example, $E_n$
for Yb in the ground configuration would be given as
\begin{equation}
  E_n = E_{6s} = E([{\rm Xe}]4f^{14}6s) - E([{\rm Xe}]4f^{14}6s^2)),
\end{equation}
with [Xe] being the Xe core and the configuration $4f^{14}6s$ representing 
the first ionization level of Yb; the level with one of the $6s$ electrons 
ripped away. In a similar manner the quantum defect for the $p$ level can be 
found by finding the difference from the $4f^{14}6s6p$ to the ionization state 
$4f^{14}6s$. Similarly, for Pb we determine $\nu_s$($\nu_p$) by considering 
ionization of the $6s^26p^2$ ground state to the $6s6p^2$($6s^26p$) state of 
Pb$^{\rm +}$, respectively.

Using the molecular orbital calculation directly to find
$\epsilon_{s,p}$ also confounds somewhat the determination of $Z_{\rm eff}$. 
If we had calculated atomic Yb, we would have $Z_{\rm eff}=1$, whereas if 
we had used Yb$^+$, we would have had $Z_{\rm eff}=2$.  Since we used neither, 
the value of $Z_{\rm eff}$ is not specified.  However, the molecular 
calculation does identify the permanent electric dipole moment of the molecule, 
as well as the bond length, via the distances $r^h$ ($r^l$) of the heavy (light)
atom from the center of mass.  Assuming charges $q^h$ and $q^l$ of the atoms, 
we can assert
\begin{eqnarray}
  d_{\rm mol} &=& r^h\,q^h+r^l\,q^l\\
  Q_{\rm mol} &=& q^h+q^l.
\end{eqnarray}
For neutral (ionic) molecules, $Q_{\rm mol}=0$($1$),  and we then solve
for the individual charges $q^h$ and $q^l$.  $Z_{\rm eff}$ on the 
heavy atom is then
\begin{equation}
  Z_{\rm eff} = q^h +1.
\end{equation}
For ThO (ThF$^{\rm +}$) we find $Z_{\rm eff} = 1.6$($2.3$), which we use in evaluating 
\eqref{relativistic_factor}.

In order to check the accuracy of this method, we applied it to several 
molecules studied previously and present the results in Table \ref{tab1}. We 
used the same {\it ab initio} calculations as in Ref. \cite{meyerbohn07}, 
but extracted the values of $\epsilon_{s,p}$ as prescribed above. 
As is evident, the new method improves our older method, yielding results within 
25\% of previously published values, thereby making it useful for estimating experimental 
feasibility of candidate molecules. The largest deviation in the table is that 
of HfF$^{\rm +}$. This deviation can be attributed mainly to the peculiar ionization 
route of Hf, which goes from $5d^26s^2$ to $5d6s^2$, indicating that the 
$d$-electron interloper plays a large role in the low-lying Hf energy levels and 
our single-channel quantum defect approach is too naive. This deficiency would 
conceivably be cleared up by a multi-channel approach to the quantum defects, which 
we have not attempted here.

\begin{table}
  \caption{\label{tab1} Comparisons of published values of ${\bf F}_{\rm eff}$ to 
    old results from Ref. \cite{meyerbohn07} and new results in the present work. 
    All values are given in GV/cm.}
  \begin{ruledtabular}
    \begin{tabular}{l|l|l|l}
      Molecule & Published & Old\cite{meyerbohn07} & New \\
      \hline
      BaF & 7.4\cite{baf} & 5.1 & 6.1\\
      \hline
      YbF & 26\cite{ybf} & 43 & 32\\
      \hline
      HgF & 99\cite{pbf} & 68 & 95\\
      \hline
      PbF & -29\cite{pbf} & -36.6 & -31\\
      \hline
      a(1) PbO & 26.2\cite{pbo} & 3.2\cite{note} & 23\\
      \hline
      HI$^{\rm +}$ & 0.34\cite{Isaev} & 0.57 & 0.34\\
      \hline
      HfF$^+$ & 24\cite{Petrov} & 18 & 30\\
      \hline
      ThO & N/A & N/A & 104\\
      \hline
      ThF$^{\rm +}$ & N/A &  N/A & 90
    \end{tabular}
  \end{ruledtabular}
\end{table}

These difficulties are not present for the Th atom. We therefore suspect that the 
predicted ${\bf F}_{\rm eff}$ in Table \ref{tab1} probably lie within 25\% 
of the answer that the fully relativistic calculations will ultimately 
provide. In more detail, to construct the molecular wave function for Th we
use the {\sc molpro} software suite\cite{MOLPRO}. The
aug-cc-pVQZ basis of Dunning\cite{avqz} is used to describe
the O and F atom's $s-f$ orbitals. The ECP78MWB of the
Stuttgart group\cite{stutt} is used to describe the 78
electron core potential of Th and the aug-cc-pVQZ basis set to describe
$s$-$f$ orbitals\cite{stutt}. We performed the calculation with an occupied 
(active) space of \{7,3,3,1\} (\{5,3,3,1\}). We took several points in $R$ 
separating Th from O or F between 2.75 and 4.5\,a$_{\rm 0}$, in order to
determine the ground state bond length of $r = 3.47$ ($3.73$)\,a$_{\rm 0}$ 
for ThO (ThF$^{\rm +}$) respectively. The value of $r$ for ThO agrees well 
with the experimental value of $r=3.48$\cite{herzberg}. The final
calculation included the present electronic state of interest ($^3\Delta$) as
well as $X\,{}^1\Sigma^+$, ${}^1\Pi$, ${}^3\Pi$, ${}^3\Sigma^+$ and ${}^1\Delta$ 
for both ThO and ThF$^{\rm +}$.

Based on this calculation we find that the effective electric field in ThO 
(ThF$^{\rm +}$)is approximately ${\bf F}_{\rm eff}\approx 104$\,GV/cm ($90$\,GV/cm). 
This similarity in ${\bf F}_{\rm eff}$ is perhaps not surprising since the 
two molecules are isoelectronic. These comparatively large values make the 
metastable ${}^3\Delta_1$ states of ThO and ThF$^{\rm +}$ competitive with 
even HgF as viable candidates for an eEDM search. There are two more parameters 
of experimental relevance that need to be estimated along with ${\bf F}_{\rm eff}$; 
the $\Omega$-doublet splitting and the lifetime of the metastable state. 

An order of magnitude estimate of the $\Omega$-doublet splitting can be 
estimated from the molecular parameters determined 
in the calculation and the use of perturbation theory\cite{brown}. 
We find that the splitting is on the order of $10^{-6}$\,cm$^{\rm -1}$ 
in both ThO and ThF$^{\rm +}$ yielding a critical field 
of a few hundred mV/cm in the metastable states, which is the field 
required to polarize the molecule.

The lifetime can be estimated by finding the amount of ${}^1\Pi_1$ that is 
admixed into the ${}^3\Delta_1$ state due to the spin-orbit interaction of 
the molecule. ${}^1\Pi_1$ can decay to the ground ${}^1\Sigma_0$ state 
through electric dipole radiation, whereas the ${}^3\Delta_1$ state would not 
decay via this route due to symmetry constraints. From our calculations, we find 
that the lifetime of the ${}^3\Delta_1$ state is on the order of $1$\,ms 
($100$\,ms)for ThO (ThF$^{\rm +}$). This is long enough to make a measurement of 
this state in a slowed molecular beam experiment\cite{acme} or in an 
ion trap\cite{eac}.

In summary, we provide a modified and more efficient method of calculating 
${\bf F}_{\rm eff}$ in polar diatomic molecules. Our determination of 
${\bf F}_{\rm eff}$ in several test cases lies within 25\% of fully relativistic 
calculations. We apply this method to the ${}^3\Delta_1$ states of 
ThO and ThF$^{\rm +}$ and find that they are extremely good candidates for 
an eEDM search yielding effective fields of $\approx104$\,GV/cm and 
$90$\,GV/cm, respectively. In addition, the $\Omega$-doublet splitting in these 
molecules is very small, allowing for easy polarization. The lifetimes of the 
metastable states are long enough to make spectroscopic measurements\cite{acme,eac}.

This work was supported by the NSF. We thank the ACME collaboration and E. A. 
Cornell for useful discussions.


\begin{thebibliography}{10}
\bibitem{note}
The original calculation in [15] did not fully account for spipn orbit effects
  which are more fully addressed in the new method.
\end{thebibliography}
\end{document}